**TITLE**

# Attosecond light field synthesis for electron motion control

**AUTHORS**

Husain Alqattan,[1] Dandan Hui,[1] Vladimir Pervak,[2] and Mohammed Th. Hassan[1,a)]

**AFFILIATIONS**

[1]Department of Physics, University of Arizona, Tucson, AZ 85721, USA.

[2]Ludwig-Maximilians-Universität München, Am Coulombwall 1, 85748, Garching, Germany.

[a)]**Author to whom correspondence should be addressed:** mohammedhassan@arizona.edu

**ABSTRACT**

The advancement of the ultrafast pulse shaping and waveform synthesis allowed to coherently control the atomic and electronic motions in matter. The temporal resolution of the waveform synthesis is inversely proportional to the broadening of its spectrum. Here, we demonstrate the light field synthesis of high-power waveforms spanning two optical octaves, from near-infrared (NIR) to deep-ultraviolet (DUV) with attosecond resolution. Moreover, we utilized the all-optical field sampling metrology for on-demand tailoring of light field waveforms to control the electron motion in matter. The demonstrated synthesis of the light field and the electron motion control pave the way for switching the photo-induced current signal in dielectric nanocircuit and establishing ultrafast photonics operating beyond the petahertz speed.

**INTRODUCTION**

Over the last few decades, different approaches have been developed to control and shape the ultrafast laser pulses with picosecond and femtosecond resolution.[1] The most common developed technique for pulse shaping is the Fourier synthesis approach (the 4f-line arrangement) which was proposed in 1983 by Froehly *et al*.[2] In this approach, the input pulse is decomposed into its constituent spectral components by a set of a grating and a lens (or curved mirror). Then, a second similar set reconstructs the output pulse. The decomposed pulse is focused— at the Fourier plane between the two sets— on a mask introduced to modulate the phase, amplitude, and/or polarization of the dispersed spectral components.[1, 3, 4] Different kinds of masks can be employed for pulse shaping such as; liquid crystal modulators,[5, 6] acousto-optic modulators,[7, 8] micro-mechanical deformable mirror,[9] and micro-electro-mechanical systems.[10] This pulse shaping approach and the temporal profile tailoring allowed coherent quantum control of matter[11, 12], and chemical reactions.[13-15]

Another pulse shaping and waveform synthesis technique can be realized by controlling the amplitude and the frequency comb's harmonics phase.[16] In this technique, the harmonics are generated by illuminating the molecular medium, such as $H_2$, with the idler and the second harmonic of a nanosecond laser pulse. The different harmonics propagate through two liquid crystal spatial light modulators to control the relative phase and amplitude of the individual harmonics. A train of synthesized waveforms is generated, by superimposing these modulated harmonics, in the shape of a square, sawtooth, sub-cycle sine and cosine, and triangular pulses.[16] The drawbacks of this technique are that the generated synthesized waveforms are in a train form, and carry low power (due to the low damage threshold of the spatial light modulators), which limits the ability to utilize these waveforms in many applications.

Alternatively, sub-cycle resolution of waveform synthesis was achieved by controlling the relative amplitudes and phases of the broad octave-spanning optical spectrum components.[17-28] This synthesis approach can be described as follows: The spectrum of a light waveform is decomposed into its constituent spectral components by a disperser, then adequate modulators act on the individual component to adjust its relative phase (or delay) and amplitude before they are coherently superimposed to obtain the desired tailored waveform. The basic principle of this synthesis approach is illustrated in Fig. 1.[1]

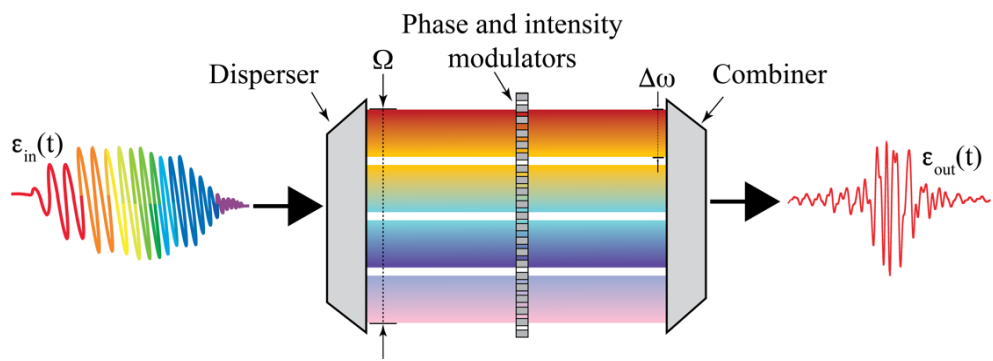

**FIG. 1.** The basic principle of light field synthesis.

Although any arbitrary field waveform can be composed by applying the above idea, in reality, the arbitrariness is sensitive to two key parameters: (i) the spectral resolution of the disperser ($\delta\omega$), which defines the temporal window (T) within which the control can be exerted, and (ii) the total spectral bandwidth ($\Omega$), which defines the shortest temporal feature ($\delta\tau$) that can be sculpted upon the intensity profile of the pulse. If the total bandwidth is narrower than the central frequency, the temporal synthesis features are encoded on the pulse's envelope over the time window (T). However, if the total bandwidth is comparable or preferably greater than the central frequency ($\Omega \geq \omega_0$), the synthesis temporal resolution enables the tailoring of the field underneath the pulse envelope.[18]

In this work, we utilized this approach to demonstrate the light field synthesis of broad bandwidth laser pulses that span over two octaves in the visible frequencies and the flanking ranges, which permits the field synthesis with attosecond resolution. This synthesis enables the sub-femtosecond electron motion control in different systems.[19, 20, 29-32]

**Generation of more than two optical octaves intense supercontinuum**

In our presented synthesis scheme, a supercontinuum coherent light is generated based on the nonlinear propagation of multi-cycle laser pulses in a hollow-core fiber[19, 20, 33]. These multi-cycle pulses, with a central wavelength ~800 nm (spectrum is shown in Fig. 2(a)), are generated from an OPCPA-based laser system (~0.75 mJ (15W), the repetition rate is 20 kHz). The temporal profile of these pulses is measured by a TG-FROG device and plotted in Fig. 2(b); the measured FROG trace is shown in Fig. 2(b) inset. These pulses have a duration of $\tau_{FWHM} \approx 15$ fs and focused to a spot size of ~170 µm by a plano-convex lens (f = 1 m) to enter a hollow-core fiber (HCF) that has a 250 µm core diameter, where the pointing of the beam is stabilized (by Aligna system from TEM Messtechnik GmbH). The input beam energy/power is in the order of ~0.5 mJ/10 W. An apparatus chamber that houses the HCF, illustrated in Fig. 2(c), is filled with Ne gas at ~2.5 bar. This chamber has entrance and exit windows of 0.5 mm thick —anti-reflection coated—UV-grade fused silica. The exit window is mounted at Brewster's angle of ~800 nm in order to provide efficient transmission over a broad spectral range. The output beam has a broadband spectrum spanning from 200–1000 nm (> 2 octaves), as shown in Fig. 2(d). This supercontinuum output beam has a power of 5 W (~50% throughput), and it is collimated using a UV-enhanced aluminium curved mirror to have a beam size of 7 mm (beam profile is shown in the inset of Fig. 2(c)). Notably, the HCF's output beam has long-time intensity stability (Root-Mean-Square (RMS) deviation value of 0.5%), which allows performing different types of experiments that last a few hours.

**High-power attosecond light field synthesizer**

The high-power supercontinuum is directed to what we so-called Attosecond Light Field Synthesizer (ALFS) device, where the light field tailoring occurs with the attosecond resolution.[20] Inside the ALFS (Fig. 3(a) and 3(b)), the supercontinuum is divided into four, nearly equal in bandwidth, constituent channels by three dichroic beam splitters (DBSs). The DBSs are designed to reflect a specific portion of the supercontinuum and transmit the rest. The four spectral channels of the synthesizer, which each spans over approximately 0.5-octave, are as follows: (i) Ch$_{NIR}$, the near-IR (700–1000 nm), (ii) Ch$_{VIS}$, the visible (500–700 nm), (iii) Ch$_{VIS-UV}$, the visible-ultraviolet (350–500 nm), and (iv) Ch$_{DUV}$, the deep ultraviolet (250–350 nm), as shown in Fig. 3(c). The pulses of the constituent channels in the synthesizer are temporally compressed by six dispersive (chirped) mirrors.[34] The temporal characterization of the individual channel's pulses is performed by a TG-FROG apparatus. The retrieved FWHM pulse durations are: $\tau_{Ch(NIR)}$ = 8.5 fs, $\tau_{Ch(VIS)}$ = 7.8 fs, $\tau_{Ch(VIS-UV)}$ = 4.5 fs, and $\tau_{Ch(DUV)}$ = 5.5 fs, as shown in Fig. 4. Remarkably, the compressed DUV pulse in the spectral region of 250–350 nm is, to our knowledge, the shortest intense pulse generated in this spectral region. This new $Ch_{DUV}$ tool would greatly impact the light field synthesis and would also be used to pump the chemical reactions in a few femtoseconds time scale.

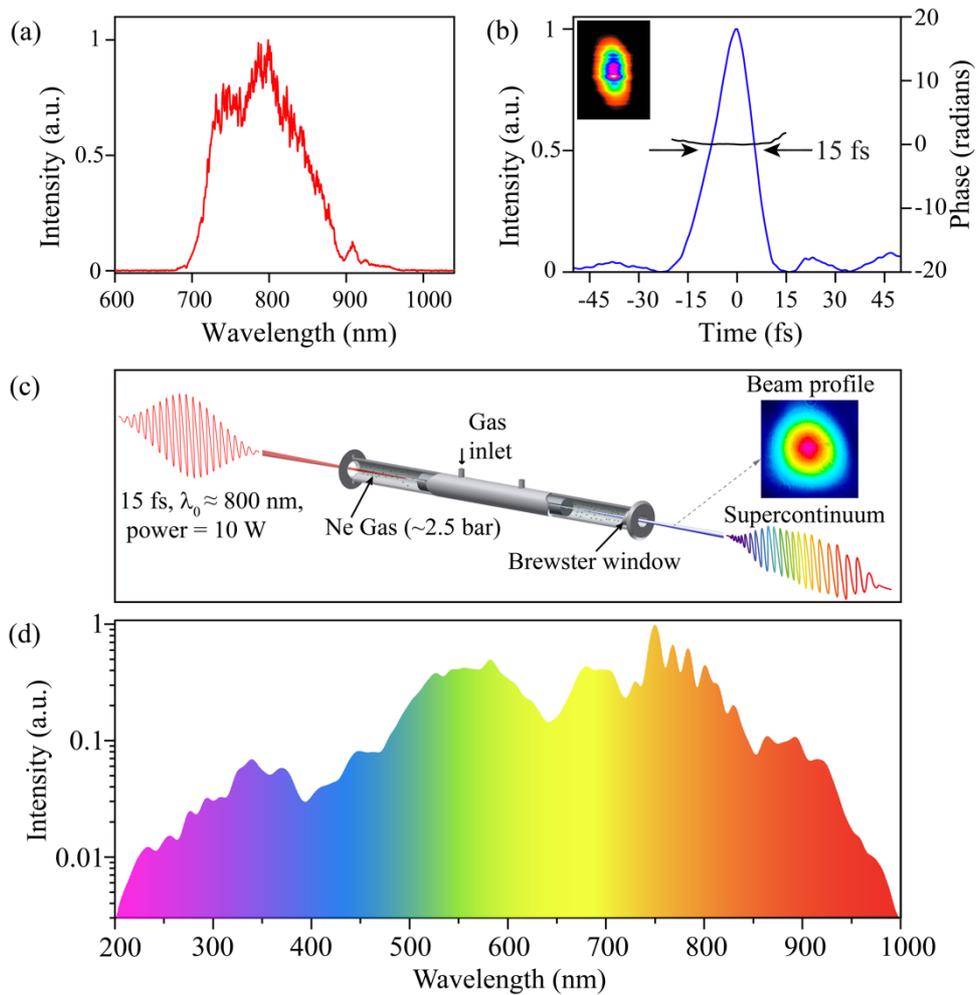

**FIG. 2.** Supercontinuum light generation spanning over more than two octaves. (a) The measured spectrum and (b) the pulse duration of the laser pulses from the source. (c) The Hollow-Core-Fiber (HCF) setup for generating the 2-octaves supercontinuum. (d) The generated supercontinuum extending over the visible and neighboring ranges (200–1000 nm), which has an energy of ~250 μJ (5W) with a homogenous gaussian distribution beam profile (shown in the inset of (c)).

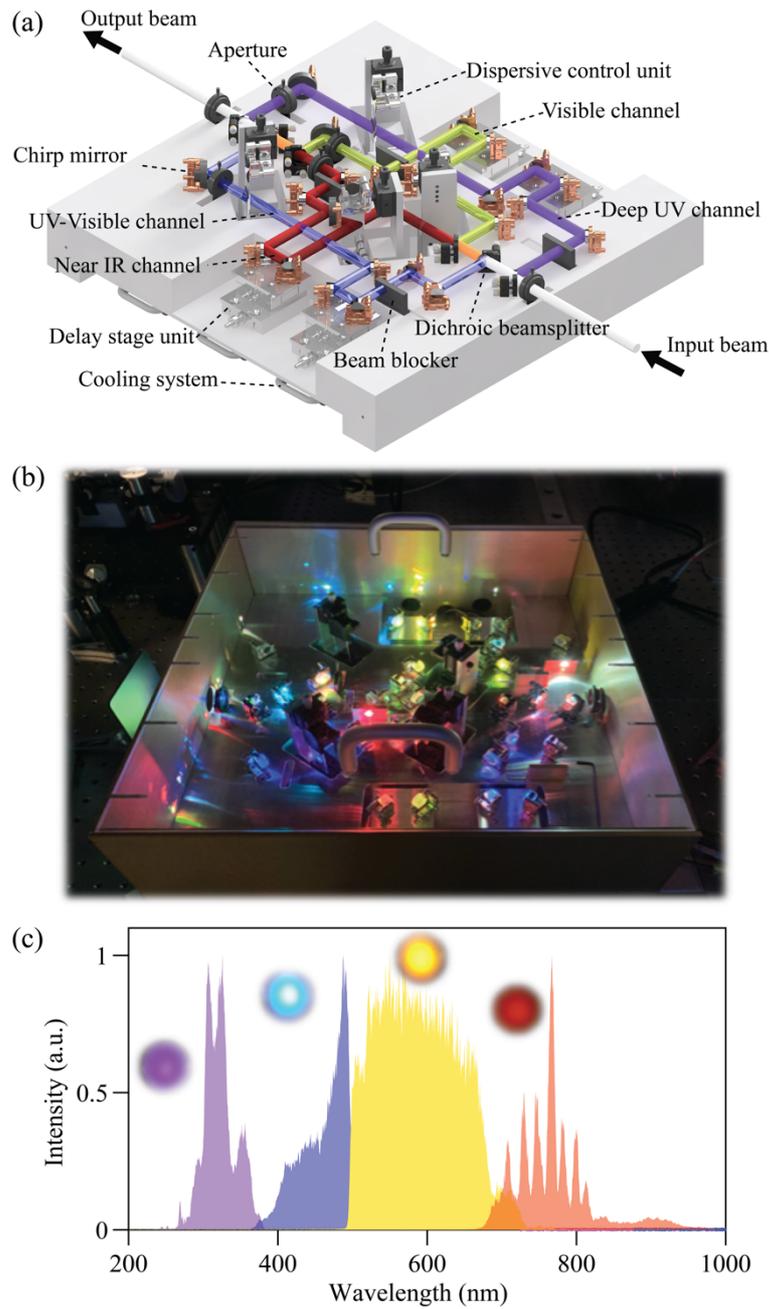

**FIG. 3.** High-Power Attosecond Light Field Synthesizer (ALFS). (a) Schematic representation of the four-channel ALFS. (b) ALFS in action. (c) The normalized spectrum of the individual channel, $Ch_{NIR}$, $Ch_{VIS}$, $Ch_{VIS-UV}$, and $Ch_{DUV}$, are shown individually in red, yellow, blue, and violet shaded curves, respectively presenting the spectral distribution in each channel.

Two elements have been implemented inside the ALFS apparatus (Fig. 3(a) and 3(b)) to synthesize the generated waveforms with attosecond resolution. These elements are (i) translation units and (ii) variable-neutral dentistry filters. The translation unit carries a pair of mirrors in the path of each constituent channel to adjust their relative path lengths (relative phase delay). It consists of a manually adjustable translation stage which is used for coarse adjustment of the optical path, and a piezoelectric translational stage, which is used for fine adjustment of the relative phases between the four channels in the desired attosecond precision. The variable neutral dentistry filters are introduced in $Ch_{NIR}$ and $Ch_{VIS}$'s beam paths to control their relative intensities. In our synthesis approach, these elements, the translation units, and the variable neutral density filters, act as the phase and amplitude modulators. At the exit of the ALFS, the constituent channel's pulses are superimposed by another set of identical DBSs. The generated synthesized waveform carries maximum power of ~2 W distributed among the four channels: $Ch_{NIR}$ ~ 1.5 W, $Ch_{VIS}$ ~ 300 mW, $Ch_{VIS-UV}$ ~ 135 mW, and $Ch_{DUV}$ ~ 65 mW. Notably, this synthesizer is providing four times higher power than the previous generations[18-20]. Moreover, in the presented work, the $Ch_{VIS-UV}$ and $Ch_{DUV}$ channels (in the UV and DUV spectral regions) have powerful shorter pulse durations.

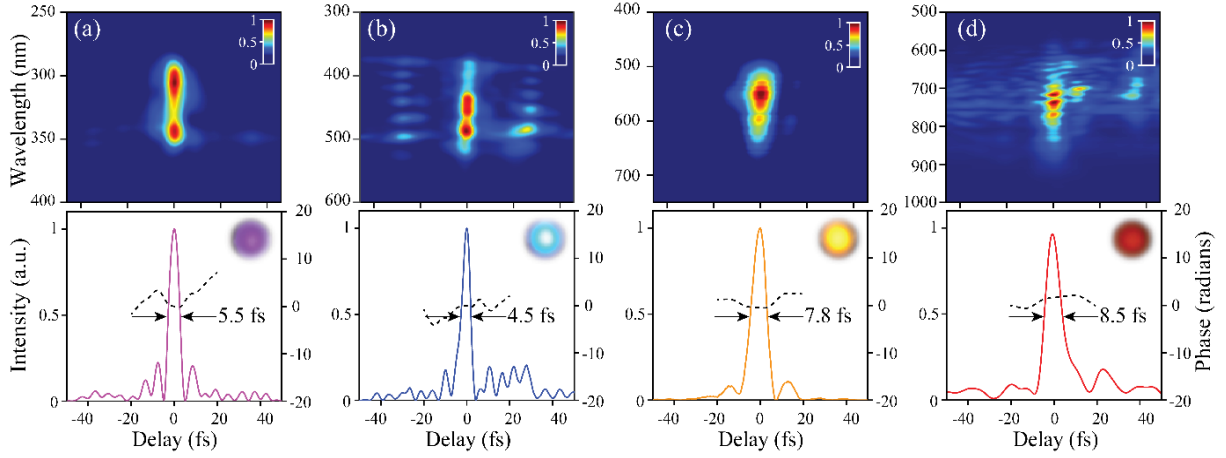

**FIG. 4.** Temporal characterization of the constituent four-channel pulses in the ALFS. The TG-FROG spectrograms are shown in the top and their retrieved temporal intensity profiles in the bottom for (a) the $Ch_{DUV}$, (b) $Ch_{VIS-UV}$, (c) $Ch_{VIS}$, and (d) $Ch_{NIR}$ pulses.

For the phase stability of this sophisticated interferometer (ALFS), the relative phases between the different pulses of the ALFS are passively and actively stabilized. The ALFS interferometer is passively thermostabilized by water cooling (at $19 \pm 0.05°C$) and enclosed in a tight housing that protects the optical setup against air fluctuations. Moreover, the ALFS is actively stabilized to compensate for any optical path drift among the four channels, based on active phase locking as demonstrated earlier[18]. In this scheme, a computer program analyzes the (few nanometers) interference spectra of the four channels' minor polarization (S-polarization) components. Accordingly, the relative phases between the s-polarized light for the four channels are determined, and the relative phases are locked at these delays. The position of the piezoelectric translational stage in each channel is actively corrected if any drift in the relative channels' delay is indicated. The typical measured RMS value for the phase stabilization between $Ch_{NIR}$ and $Ch_{VIS}$ is ~74 mrad, $Ch_{UV-Vis}$ and $Ch_{VIS}$ is ~68 mrad, $Ch_{UV-Vis}$ and $Ch_{DUV}$ is ~33 mrad.

## All-optical field sampling of synthesized waveforms

Sampling the synthesized waveforms is crucial for the on-demand light field synthesis. The light field metrology approaches, such as attosecond streaking[35] and Attoclock[36], require a fully equipped XUV beamline that includes the generation of XUV pulses by High Harmonic Generation (HHG), attosecond XUV pulse isolation, and a sophisticated and complex laser-pump XUV-probe setup.[37] Such beamline is costly and technically complicated. These technical limitations in light field sampling hold back the advancement in the ultrafast light field synthesis. An optoelectronic methodology has been demonstrated to sample the field with an accuracy comparable to the attosecond streaking technique.[38-41] This promising technique demands certain technological development in preparing the whole electronic circuit, signal detection, and the sequence current measurement tools.

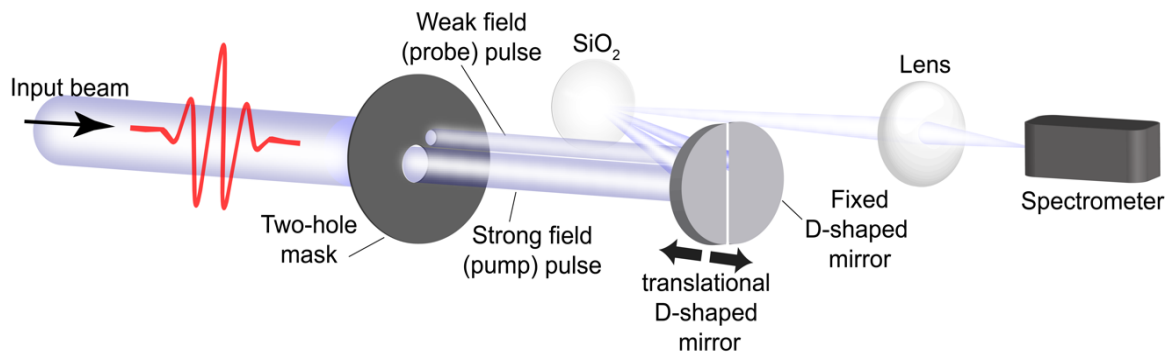

**FIG. 5.** Illustration of the all-optical field sampling setup.

Alternatively, we demonstrated a new all-optical light field metrology for sampling synthesized light fields based on the strong field-induced phase transition in a $SiO_2$[42]. This field sampling approach can be described as follows: A beam of an unknown field is split into two beams. The first beam (strong field) is intense enough to alter the optical properties and change the reflectivity of the dielectric. The second beam is set at low intensity to avoid inducing any change in the dielectric and acts as a probe beam. The photo-induced modulation in this optical reflectivity of the dielectric follows the vector potential of the unknown strong (pump) driver field. Thus, the unknown field can be retrieved by recording the intensity change of the reflected weak probe beam as a function of the delay between the two beams. We utilized this all-optical field sampling approach (setup is shown in Fig. 5), to characterize the waveforms synthesized by the ALFS. The output beam from the ALFS passes through a two-hole mask with two different hole diameters to emerge two beams with different intensities (the pump and probe). The two beams propagate and focus onto a 100 μm dielectric (SiO₂) sample by two separate D-shaped focusing mirrors (f = 100 mm). One of these mirrors is attached to a nanopositioning stage to control the time delay between the two beams with attosecond resolution. The spectrum of the reflected probe beam of the SiO₂ sample is recorded, by the spectrometer, as a function of this time delay (with steps of 100 as). The Total Reflectivity Modulation (TRM) trace is obtained by integrating the reflected spectrum at each time instance identical to the vector potential of the measured unknown waveform[42]. The electric field of the strong driver pulses can be retrieved by calculating the time derivative of the vector potential (obtained from the TRM):

$$E(t) = \frac{\partial A(t)}{\partial t}. \tag{1}$$

The derivation is carried out in the frequency domain on $A(\omega)$, which is the Fourier transform of $A(t)$. To avoid the high-frequency noise, a bandpass filter is applied, $H(\omega)$, as shown in Eqs. (2) and (3). The spectral window can be specified by changing its width, $\sigma$, and center, $\omega_0$, based on the pump pulses' measured spectrum.

$$E(t) = \int_{-\infty}^{\infty} i\omega \times A(\omega) \times H(\omega) \times e^{i\omega t} \, d\omega, \tag{2}$$

where,

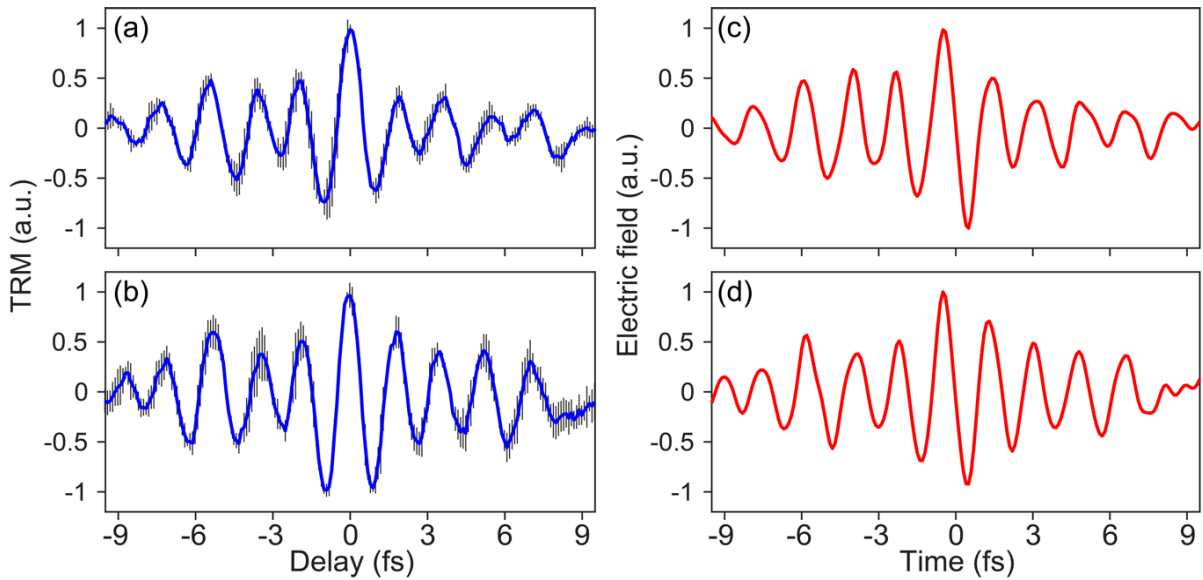

**FIG. 6.** All-optical light field sampling of waveforms generated by ALFS: (a) and (b) The measured TRMs. (c) and (d) The retrieved electric fields of synthesized waveforms.

$$H(\omega) = \exp\left(-\frac{1}{2}\left(\frac{\omega-\omega_0}{\sigma}\right)^{order}\right). \tag{3}$$

Some of the sampled synthesized waveforms are shown in Fig. 6. The measured TRM (average of three measurements) is shown in the left column, and the corresponding retrieved fields are shown in the right column.

**Attosecond laser pulses (ALPs)**

The synthesis of the half-cycle laser pulse with a duration < 1fs by the ALFS — henceforth, it is called attosecond laser pulse (ALP)— is possible by spatiotemporally overlapping of all the four channels in the ALFS and modifying the relative phases and intensities of the ALFS channels. The desired relative amplitudes ratio is achieved by adjusting the variable neutral dentistry filters in Ch$_{NIR}$ and Ch$_{VIS}$. It took a few iterations to fine adjust the relative delays between the four channels to achieve the shortest ALP, which is sampled and the TRM is shown in Fig. 7(a). The ALP's retrieved field and its instantaneous intensity are shown in Fig. 7(b) and 7(c), respectively. The ALP is carried at central wavelength $\lambda_0$ ~545 nm and has an energy of ~20 µJ. The FWHM temporal duration for the central field crest of the ALP is ~400as.

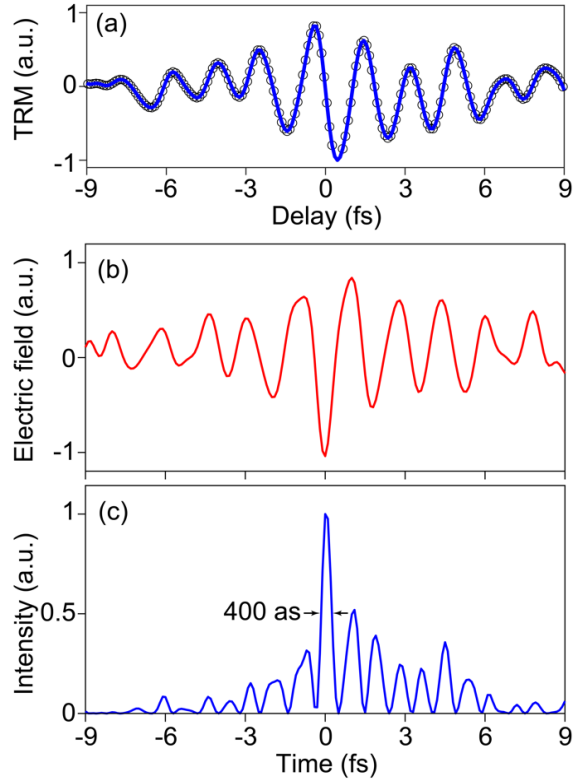

**FIG. 7.** Attosecond laser pulse (ALP): (a) The measured TRM, (b) retrieved electric field, and (c) the instantaneous intensity. The temporal duration of the main crest is $\tau_{FWHM} \approx 400$ as.

## The light field synthesis scheme

The all-optical field sampling characterization of the synthesized waveform and the full control of pulse intensities and relative delays for the four channels from the ALFS enables the on-demand light field synthesis, which was not simply before since the waveform sampling was done using the complex XUV attosecond streaking technique[18-20].

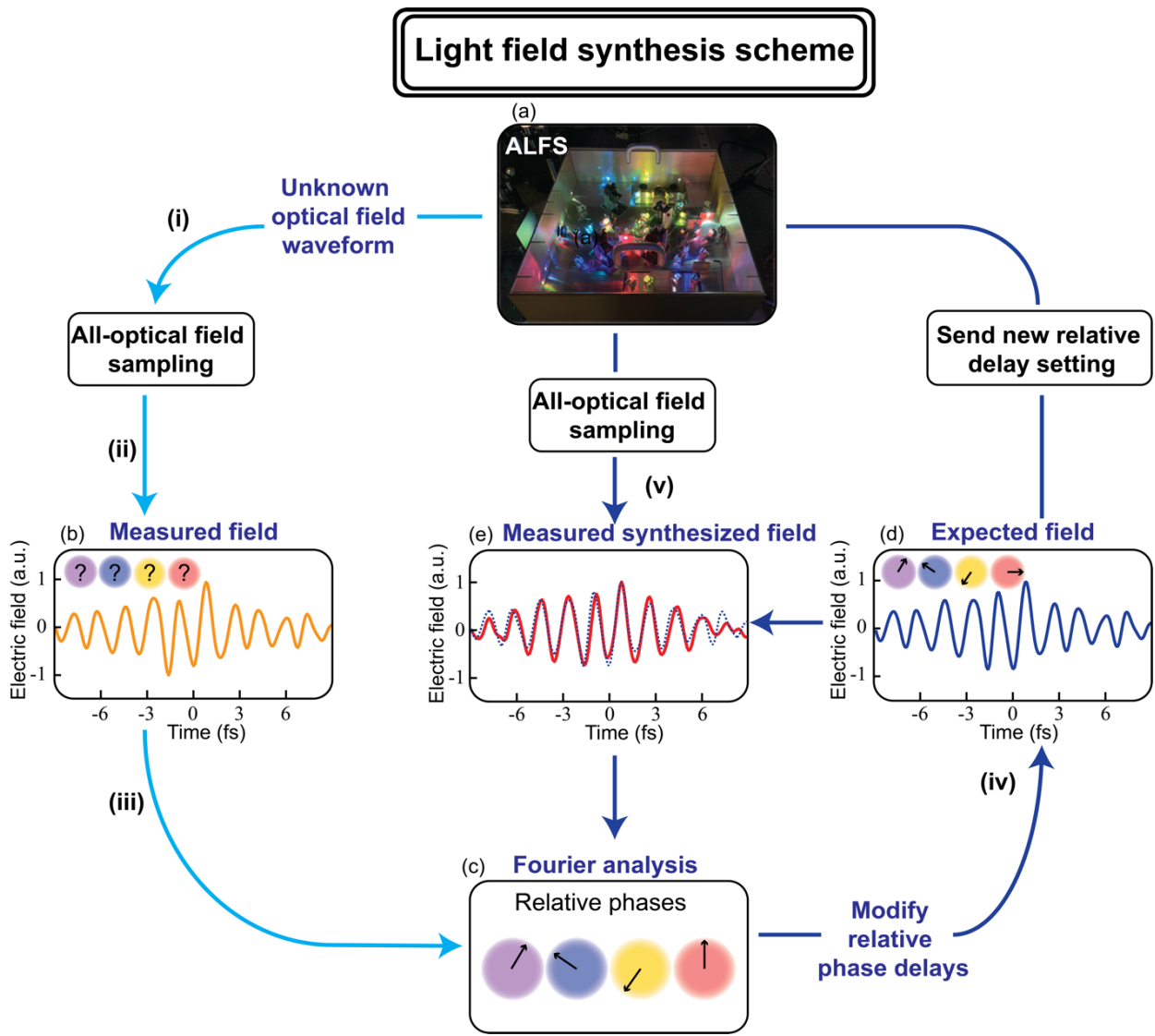

**FIG. 8.** The on-demand light field synthesis scheme. (a) The unknown waveform is generated by ALFS. The field synthesis process occurs in the following steps: (i) the unknown field is sampled by the all-optical field sampling method. (ii) the measured field is retrieved and plotted in (b). (iii) the relative phases of the ALFS channels are determined by the Fourier transforming of the measured waveform as shown in (c). (iv) the relative phase between the channels is modified to create the desired waveform by calculation in (d); then, the new delay setting between the channels is sent to the delay stages inside the ALFS. (v) sampling the new generated synthesized waveform, which is plotted in comparison with the expected waveform in (e). This synthesis process is repeated to generate, on-demand, other desired waveforms.

Accordingly, we developed a light field synthesis scheme (illustration is shown in Fig. 8), which can be summarized in the following steps:

(i) A synthesized waveform is generated by the ALFS (Fig. 8(a)) at which the four channels have certain unknown relative phase delay setting. (ii) The vector potential and the retrieved electric field are measured (Fig. 8(b)) using the all-optical field sampling approach described earlier. (iii) The relative phases between the four spectral channels' pulses are extracted by Fourier transforming the unknown synthesized waveform (Fig. 8(c)). Then the phases of the ALFS channels are actively locked. (iv) A desired simulated synthesized waveform is created by changing the relative phase delays between the four channels (for example, in Fig. 8(d), $Ch_{NIR}$ is moved by -2 fs); the desired waveform is generated by sending the new relative delay setting to the nanopositioning stages in the ALFS. (v) Finally, the new measured synthesized waveform (Fig. 8(e) solid red line) is plotted in contrast with the simulated desired field (Fig. 8(d)). The two fields are found to be in profound agreement (the RMS is ~8%), demonstrating the viability of the presented light field synthesis scheme. This synthesis scheme opens the door to implement the machine learning technology in ultrafast light field synthesis and pulse shaping advancements.

This demonstrated capability of the light field synthesis allows to control the electron motion —driven by these synthesized fields—in different systems. For instance, the electron motion can be controlled on-demand in fused silica dielectric using the waveforms in Fig. 8(b) and Fig. 8(e) (in solid red line). The instantaneous field intensities of these fields are shown in

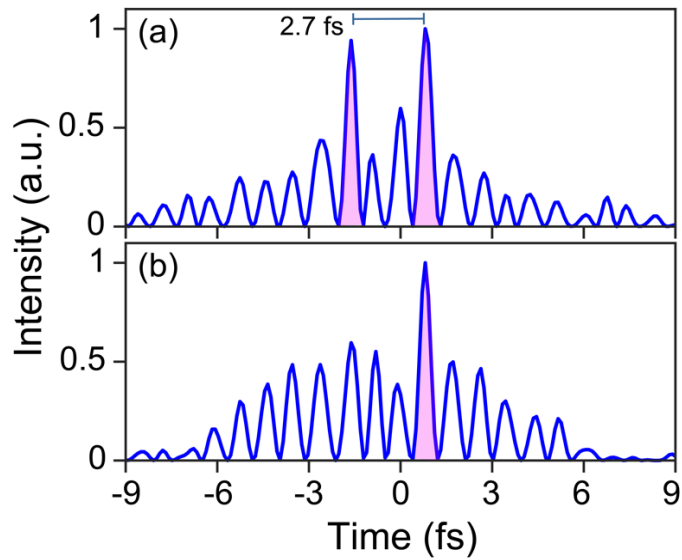

**FIG. 9.** (a) The measured field intensity of the waveform (shown in Fig. 8(b)) maximizing the number of excited electrons in fused silica at two-time instances separated by 2.7 fs. (b) The measured field intensity of the waveform that is shown in Fig. 8(e). The maximum electron excitation is confined to one field crest lasting < 1fs.

Fig. 9(a) and 9(b), respectively. These synthesized waveforms drive electron motion in the dielectric conduction band, which follows the shape of the vector potential of these fields[42]. Therefore, the maximum electron triggering occurs at the highest field amplitude. Accordingly, the synthesized field shown in Fig. 8(b) maximizes the number of triggered electrons at the highest two field crests, separated by 2.7 fs as shown by the violet shaded area in field intensity in Fig. 9(a). The excitation maximized at only one field crest (Fig. 9(b)) by the synthesized waveform in Fig. 8(e).

**CONCLUSION**

In this work, the attosecond synthesis of powerful light field waveforms— spanning two octaves from near-infrared to deep ultraviolet— is achieved using the attosecond light field synthesizer (ALFS). The capability of compressing and synthesizing such a coherent broadband light enables the generation of the attosecond laser pulses, which have a sub-femtosecond temporal duration, and tailoring the field with attosecond resolution. The synthesized waveforms were sampled by the all-optical light field sampling metrology. Moreover, we present the light field synthesis scheme, which allows on-demand control of the electron motion in a dielectric. The demonstrated work enables the control of photo-induced current signals in potential dielectric nanocircuits. This advancement opens the door to establish ultrafast electronics that promise an increase in the speed of data processing and information encoding at rates exceeding 1 Pbit/sec, extending the frontiers of modern electronics and information processing technologies.

## ACKNOWLEDGMENTS

This work partially supported by the Air Force Office of Scientific Research under award number FA9550-19-1-0025. This project is partially funded by the Gordon and Betty Moore Foundation Grant number GBMF7938. We also are grateful for The W.M. Keck Foundation for supporting this project.